\documentclass[12pt]{iopart}

\usepackage[lmargin=2.5cm,rmargin=2.5cm,tmargin=3cm,bmargin=2cm]{geometry}
\usepackage{amssymb}
\usepackage{graphicx}
\usepackage{graphics}
\usepackage{appendix}
\usepackage{multirow}
\usepackage{hhline}
\usepackage[normalem]{ulem}
\usepackage[usenames, dvipsnames]{color}
\usepackage{relsize}
\usepackage{color}
\usepackage[font=small]{caption}
\usepackage{subcaption}
\usepackage{xspace}
\usepackage{cite}

\usepackage{hyperref}
\hypersetup{
    colorlinks=true,
    linkcolor=blue,
    filecolor=blue,      
    urlcolor=blue,
    citecolor=blue
}

\usepackage[capitalize]{cleveref}

\newcommand{\blue}[1]{\textcolor{red}{#1}}
\renewcommand{\blue}[1]{#1}
\newcommand{\jm}[1]{\textcolor{red}{#1}}
\renewcommand{\jm}[1]{#1}
\newcommand{\mysubsubsection}[1]{\subsubsection{#1}\hfill}

\newcommand{\ppcf}[1]{\textcolor{red}{#1}}
\renewcommand{\ppcf}[1]{#1}
\newcommand{\Ndata}{\ppcf{4768}}

\newcommand{\Alfven}{Alfv\'{e}n\xspace}

\newcommand{\AlfvenEigenmode}{\Alfven Eigenmode\xspace}
\newcommand{\AlfvenEigenmodes}{\Alfven Eigenmodes\xspace}
\renewcommand{\AE}{AE\xspace}
\newcommand{\AEs}{AEs\xspace}
\newcommand{\TAE}{TAE\xspace}
\newcommand{\TAEs}{TAEs\xspace}
\newcommand{\GAE}{GAE\xspace}
\newcommand{\GAEs}{GAEs\xspace}
\newcommand{\AEADiagnostic}{\AlfvenEigenmode Active Diagnostic\xspace}
\newcommand{\AEAD}{AEAD\xspace}
\newcommand{\AEAntenna}{AE antenna\xspace}

\newcommand{\xpoint}{X-point\xspace}
\renewcommand{\etal}{\emph{et al}\xspace}
\renewcommand{\NF}{\emph{Nucl. Fusion}\xspace}
\newcommand{\EP}{EP\xspace}

\newcommand{\RF}{RF\xspace}
\newcommand{\deuterium}{deuterium\xspace}
\newcommand{\hydrogen}{hydrogen\xspace}

\renewcommand{\H}{\mathrm{H}}

\newcommand{\MISHKA}{MISHKA\xspace}
\newcommand{\EFIT}{EFIT\xspace}
\newcommand{\HELENA}{HELENA\xspace}
\newcommand{\CSCAS}{CSCAS\xspace}
\newcommand{\pdf}{pdf\xspace}
\newcommand{\pdfs}{pdfs\xspace}
\newcommand{\SparSpec}{SparSpec\xspace}

\newcommand{\wo}{\omega_0}
\newcommand{\f}{f}
\newcommand{\fo}{f_0}
\newcommand{\go}{\gamma/\wo}
\newcommand{\dgo}{\Delta(\gamma/\wo)}
\newcommand{\n}{n}
\newcommand{\absn}{\vert\n\vert}
\newcommand{\qnf}{q_{95}}
\newcommand{\Pnbi}{P_{\mathrm{NBI}}}
\newcommand{\Prf}{P_{\mathrm{RF}}}
\newcommand{\Nbin}{\mathrm{N_{bin}}}
\newcommand{\Ntot}{\mathrm{N_{tot}}}
\newcommand{\ftae}{\f_\mathrm{TAE}}
\newcommand{\fmishka}{\f_\mathrm{\MISHKA}}
\newcommand{\Bo}{B_0}
\newcommand{\Ro}{R_0}
\newcommand{\uo}{\mu_0}

\newcommand{\mH}{m_\mathrm{H}}
\newcommand{\qo}{q_0}
\renewcommand{\ne}{n_{\mathrm{e}}}
\newcommand{\neo}{n_{\mathrm{e}0}}

\newcommand{\Teo}{T_{\mathrm{e}0}}

\newcommand{\meff}{m_\mathrm{eff}}
\newcommand{\Rsquare}{R^2}
\newcommand{\Iant}{I_\mathrm{ant}}
\newcommand{\chisquare}{\chi^2}
\newcommand{\Np}{N}
\newcommand{\nmax}{\n_\mathrm{max}}
\newcommand{\nres}{\n_0}
\newcommand{\nant}{\n_\mathrm{ant}}
\newcommand{\cf}{X}
\newcommand{\cfSS}{A}
\newcommand{\Ip}{I_\mathrm{p}}
\newcommand{\dlfs}{d_\mathrm{sep}}
\newcommand{\dlcfs}{\dlfs}
\newcommand{\snf}{s_{95}}
\newcommand{\elon}{\kappa}
\newcommand{\triu}{\delta_u}
\newcommand{\tril}{\delta_l}
\newcommand{\betaN}{\beta_\mathrm{N}}
\newcommand{\li}{\ell_\mathrm{i}}
\newcommand{\felm}{f_\mathrm{ELM}}

\newcommand{\halfwidth}{0.49\columnwidth}

\newcommand{\SI}[2]{#1~\mathrm{#2}}
\newcommand{\rd}{\mathrm{d}}

\newcommand{\JPN}[1]{\mathrm{JPN}~#1}
\newcommand{\N}[1]{(\Ntot = #1)}
\newcommand{\abs}[1]{\vert #1 \vert}
\newcommand{\norm}[1]{\vert\vert #1 \vert\vert}

\newcommand{\Ni}{\n_i}
\newcommand{\Nj}{\n_j}

\newcommand{\pk}{\phi_k}
\newcommand{\ti}{\theta_i}
\newcommand{\tj}{\theta_j}
\newcommand{\tk}{\theta_k}
\newcommand{\Nmax}{\nmax}
\newcommand{\Nstar}{\n^*}
\newcommand{\mod}[1]{\mathrm{mod}\left( #1,2\pi\right)}
\newcommand{\atantwo}[1]{\mathrm{atan2}\left(#1\right)}
\renewcommand{\xi}{x_i}
\newcommand{\yi}{y_i}
\newcommand{\xj}{x_j}
\newcommand{\yj}{y_j}
\newcommand{\floor}[1]{\lfloor#1\rfloor}
\newcommand{\mk}{m_k}

\renewcommand{\vec}[1]{\mathbf{#1}}
\newcommand{\vphi}{\vec{\phi}}
\newcommand{\vx}{\vec{x}}
\newcommand{\vy}{\vec{y}}

\newcommand{\iPSFC}{$^1$\xspace}
\newcommand{\iEPFL}{$^2$\xspace}
\newcommand{\iCCFE}{$^3$\xspace}
\newcommand{\iJET}{*\xspace}

\newcommand{\PSFC}{\iPSFC Plasma Science and Fusion Center, Massachusetts Institute of Technology, Cambridge, MA, USA\xspace}
\newcommand{\EPFL}{\iEPFL Swiss Plasma Center, Ecole Polytechnique F\'{e}d\'{e}rale de Lausanne, Lausanne, Switzerland} 
\newcommand{\CCFE}{\iCCFE Culham Centre for Fusion Energy, Culham Science Centre, Abingdon, UK} 
\newcommand{\JET}{\iJET See author list of E. Joffrin \etal 2019 \NF 59 112021\xspace}

\begin{document}

    \title[]{Results from the \AEADiagnostic during the 2019-2020 JET \deuterium campaign}
    
    \author{R.A.~Tinguely\iPSFC\footnote{Author to whom correspondence should be addressed: rating@mit.edu}, 
    P.G.~Puglia\iEPFL, 
    N.~Fil\iCCFE, 
    S.~Dowson\iCCFE, 
    M.~Porkolab\iPSFC, 
    A.~Fasoli\iEPFL, 
    D.~Testa\iEPFL, 
    and JET~Contributors\iJET}
    
    \address{\PSFC \\
             \EPFL \\
             \CCFE \\
             \JET}
    
    \begin{abstract}
        This paper \blue{presents results of extensive analysis of mode excitation observed during} the operation of the \AEADiagnostic (\AEAD) in the JET tokamak during the 2019-2020 \deuterium campaign. Six of eight toroidally spaced antennas, each with independent power and phasing, were successful in actively exciting \emph{stable} MHD modes in \blue{479} plasmas. In total, \blue{\Ndata} magnetic resonances were detected with up to fourteen fast magnetic probes. In this work, we \blue{present} the calculations of resonant frequencies $\fo$, damping rates $\gamma < 0$, and toroidal mode numbers $\n$, \blue{spanning the parameter range} $\fo \approx \SI{30-250}{kHz}$, $-\gamma \approx \blue{\SI{0-13}{kHz}}$, and $\absn \leq \blue{30}$. \blue{In general, good agreement is seen between the resonant and the calculated toroidal \AlfvenEigenmode frequencies, and between the toroidal mode numbers applied by the \AEAD and estimated of the excited resonances.} We note several trends in the database: the probability of resonance detection decreases with plasma current and external heating power; the normalized damping rate increases with edge safety factor but decreases with external heating. \blue{These results provide key information to prepare future experimental campaigns and to better understand the physics of excitation and damping of \AlfvenEigenmodes in the presence of alpha particles during the upcoming DT campaign, thereby extrapolating with confidence to future tokamaks.}
    \end{abstract}
    



\section{Introduction}\label{sec:intro}

    In tokamaks, an energetic particle (\EP) population, such as radio frequency (\RF) heated ions or DT alphas, can destabilize \AlfvenEigenmodes (\AEs). In turn, these \AEs can lead to an increase in \EP transport and decrease in fusion performance. Understanding \AE stability, i.e. driving and damping mechanisms, is therefore essential to the operation and success of future tokamaks with significant alpha particle populations, such as ITER \cite{iter_Fasoli2007}, SPARC \cite{Greenwald2018}, and other devices.

    \jm{In the JET tokamak, fast ion populations, such as those resulting from ion cyclotron resonance heating, can destabilize Toroidal \AlfvenEigenmodes (\TAEs).}
    These unstable modes are typically easily identifiable as coherent structures, with well-defined frequencies and toroidal mode numbers, in the \blue{Fourier} spectra of fast magnetic probe \blue{data}. For unstable \AEs, their total growth rate is positive, $\gamma > 0$, as the fast ion drive overcomes various damping mechanisms, e.g. continuum, radiative, and electron/ion Landau damping. However, in the case of overwhelming damping, \AEs cannot be seen in the magnetic spectra without external excitation; this scenario may even occur in upcoming JET DT experiments if the alpha particle population is insufficient to destabilize the modes. Thus, in order to better study and understand \AE stability, the \AEADiagnostic (\AEAD, or \AE antenna) \cite{Fasoli1995,Puglia2016} is used in JET to actively excite, or probe, \emph{stable} \AEs and measure their resonant frequencies $\wo = 2\pi\fo$, \blue{toroidal} mode numbers $\n$, and \emph{total} damping rates $\gamma < 0$.
    
    In this paper, we provide an overview of the operation and measurements of the \AEAntenna during the 2019-2020 JET \deuterium campaign. We note that many past works have analyzed or reported data from previous campaigns; these include studies with the original, low-$\n$ saddle coils \cite{Fasoli1995,Fasoli1995nf,Fasoli1996,Fasoli1997,Heidbrink1997,Jaun1998,Wong1999,Fasoli2000,Fasoli2000pla,Jaun2001,Testa2001,Fasoli2002,Testa2003,Testa2003NBI,Testa2003rsi,Testa2004,Testa2005,Testa2006,Fasoli2007,Klein2008,Fasoli2010}, the intermediate-to-high-$\n$, eight antenna system \cite{Panis2010,Testa2010,Testa2010epl,Testa2011,Testa2011fed,Panis2012a,Panis2012b,Testa2012,Testa2014}, and the most recent upgrade \cite{Puglia2016,Nabais2018,Aslanyan2019}, among others. 
    \jm{Novel to this work are the following: (i)~The recently upgraded independent phasing of the eight \AE antennas allows probing of high toroidal mode numbers $\absn \leq 20$. (ii)~An updated magnetics system, with fourteen fast magnetic probes, allows confident measurements of $\fo$, $\gamma$, and $\absn \leq 30$. (iii)~A database of $\sim$5000~resonances are detected in $\sim$500~plasmas spanning a wide parameter space; important trends are observed in the bulk data, and identification of individual pulses opens opportunities for further study and comparison with simulation. These analyses are necessary for assessing \AE drive and damping mechanisms, validating modeling efforts, and extrapolating the impact of \AEs in future tokamaks.}
    
    The outline of the rest of the paper is as follows: In \cref{sec:AEAD}, we briefly review active excitation of \Alfven modes with the \AEAntenna. \Cref{sec:detection} describes resonance detection with the fast magnetics system and details the calculations of $\fo$, $\go$, and $\n$. In \cref{sec:opspace}, we further explore operational and parameter spaces, noting trends in the data and suggesting opportunities for further analysis. Finally, a summary is provided in \cref{sec:summary}. 


\section{Active antenna excitation}\label{sec:AEAD}

    The original \AEAntenna system consisted of saddle coils capable of exciting \AEs with low toroidal mode numbers $\absn \leq 2$ \cite{Fasoli1995,Fasoli1996}. From 2007-2008, an upgrade \cite{Fasoli2003iaea,Testa2004soft,Panis2010} involved the installation of eight in-vessel, toroidally spaced antennas - two sets of four - situated below the midplane at $R \approx \SI{3.68}{m}$, $Z \approx \SI{-0.65}{m}$ and with toroidal positions $\phi~\approx~\SI{\{0,4.7,9.4,14.1,180,184.7,189.4,194.1\}}{degrees}$. Each antenna comprises $\SI{18}{turns}$ and has poloidal and toroidal dimensions $\sim \SI{20}{cm} \times \SI{20}{cm}$. The antennas can be operated in three frequency ranges $\f = \SI{25-50}{kHz}, \SI{75-150}{kHz}$, and $\SI{125-250}{kHz}$, with each frequency filter allowing antenna currents up to $\Iant \approx \SI{10}{A}, \SI{7}{A}$, and $\SI{4}{A}$, respectively, at the maximum frequencies. A synchronous detection system is used to identify \cite{Fasoli1995nf} and track \cite{Fasoli1997} resonances in real time. 
    
\begin{figure}[h!]
            \centering
            \begin{subfigure}{\halfwidth}
                \includegraphics[width=\textwidth]{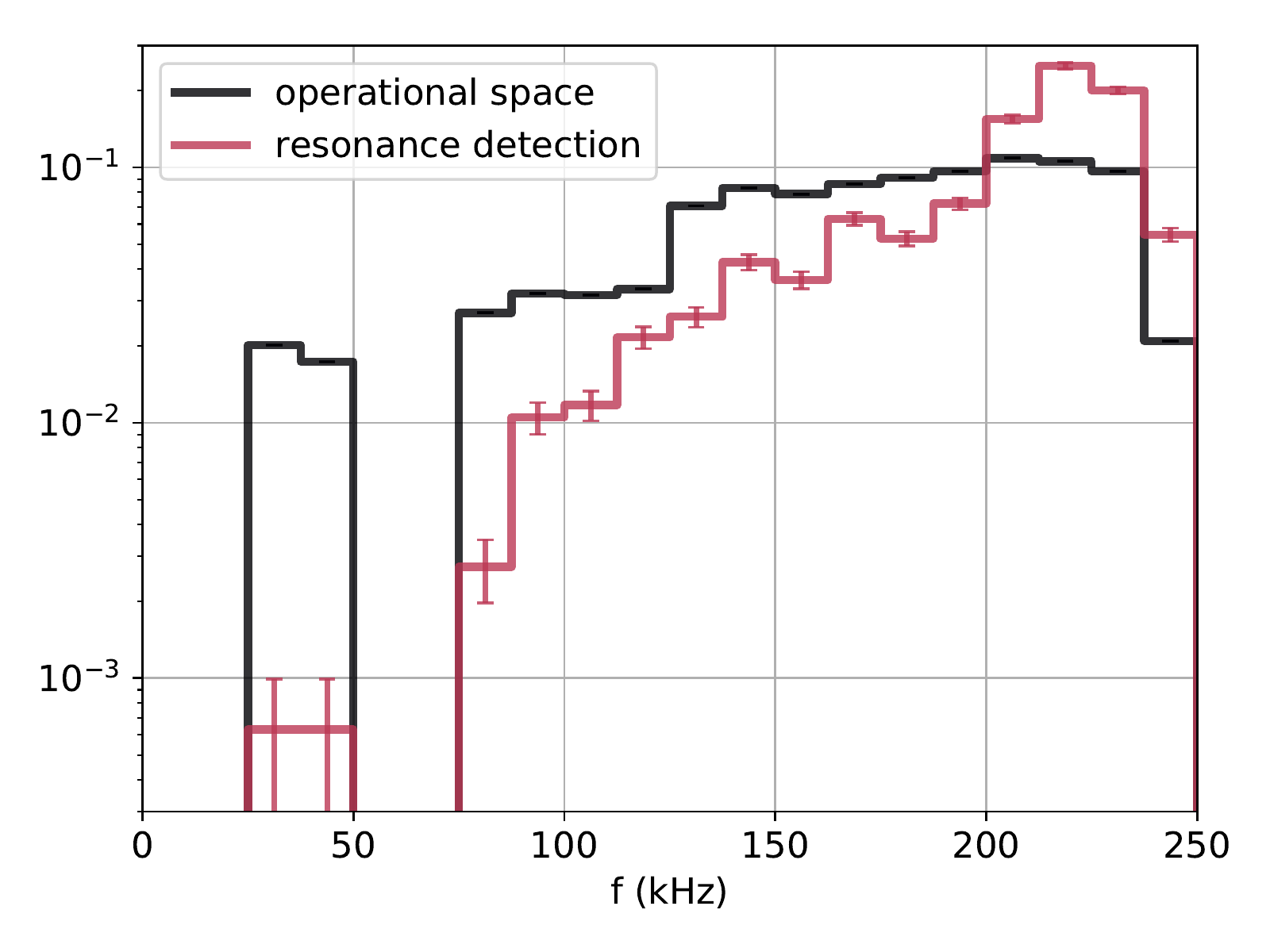}
                \caption{}
                \label{fig:fopspace}
            \end{subfigure}
            \begin{subfigure}{\halfwidth}
                \includegraphics[width=\textwidth]{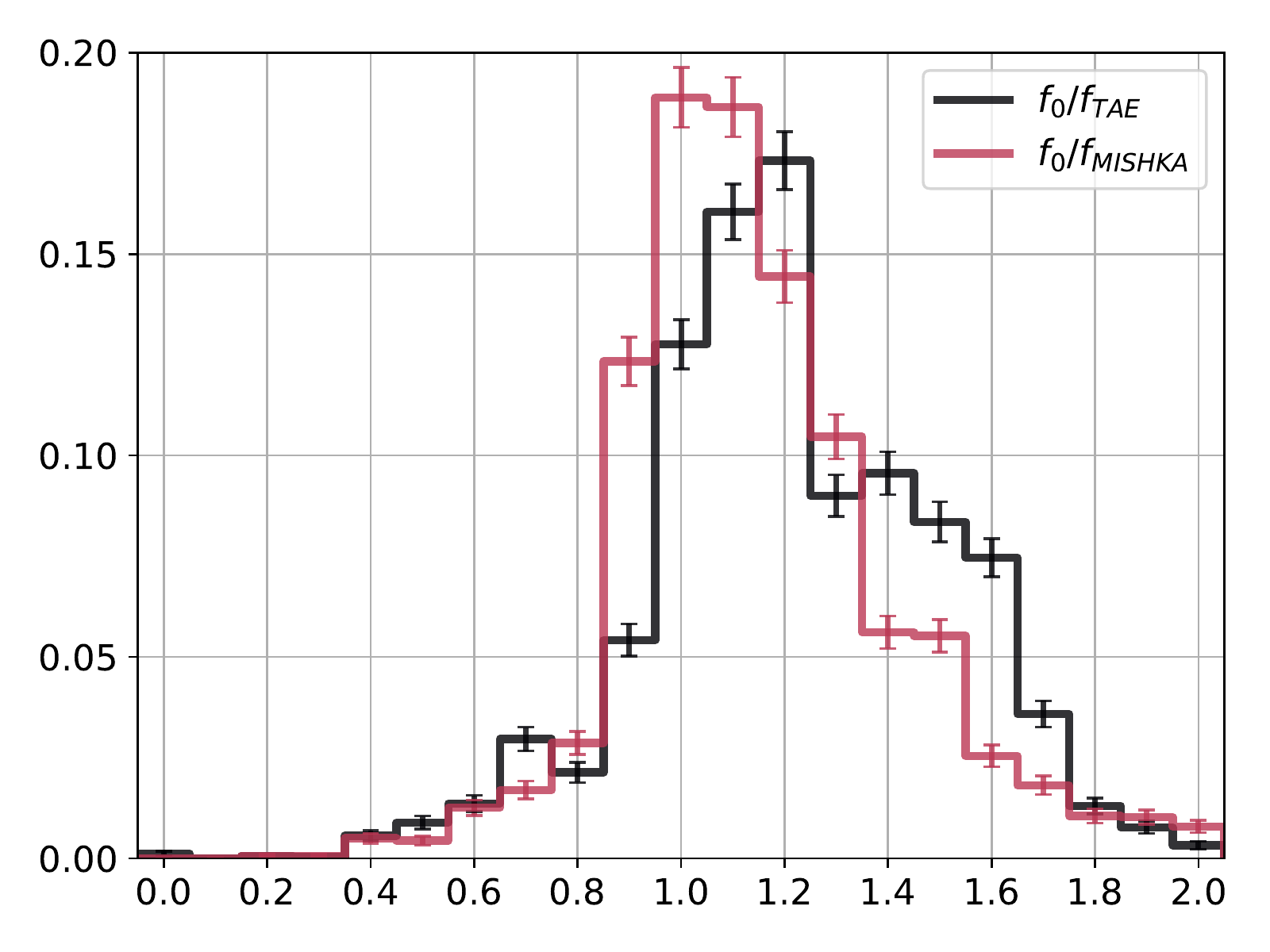}
                \caption{}
                \label{fig:fratio} 
            \end{subfigure}
            \caption{
            \jm{Histograms normalized to their total number of data points $\Ntot$: (a) The antenna operational space $(\Ntot \approx 5\times 10^6)$ and resonance detection space $\N{\Ndata}$ versus frequency. (Note the logarithmic scale of the vertical axis.) (b) Ratios comparing resonant, estimated \TAE, and \MISHKA-evaluated \cite{mishka_mikhailovskii1997} frequencies \blue{$\N{3780}$}. Uncertainties are shown as error bars.}
            }
        \end{figure}

        To find stable \AEs, the antennas' frequencies are simultaneously scanned within a given filter's range at rates $\vert \rd f/ \rd t \vert \leq \SI{50}{kHz/s}, \SI{100}{kHz/s}$, and $\SI{200}{kHz/s}$, respectively, for the filters above. The operational space of the \AEAntenna during the 2019 JET \deuterium campaign is visualized in \cref{fig:fopspace}; the histogram (black) shows the number of data points $\Nbin$ collected within each frequency bin, normalized to the total number of data points $\Ntot$. \blue{(Throughout the paper, $\Ntot$ will be noted for each histogram or distribution.) Error bars - though impossible to see in the black histogram - are included to indicate the uncertainty from counting statistics, calculated here as $\sqrt{\Nbin}/\Ntot$.} As can be seen, the system was operated more frequently with the high frequency filters, and no data exists in the inaccessible range $\f = \SI{50-75}{kHz}$. In total, the \AEAntenna was operated during 676 plasma discharges \blue{during the 2019-2020 deuterium campaign,} spanning $\JPN{93063-96855}$.
    
\begin{figure}[h!]
            \centering
            \begin{subfigure}{\halfwidth}
                \includegraphics[width=\textwidth]{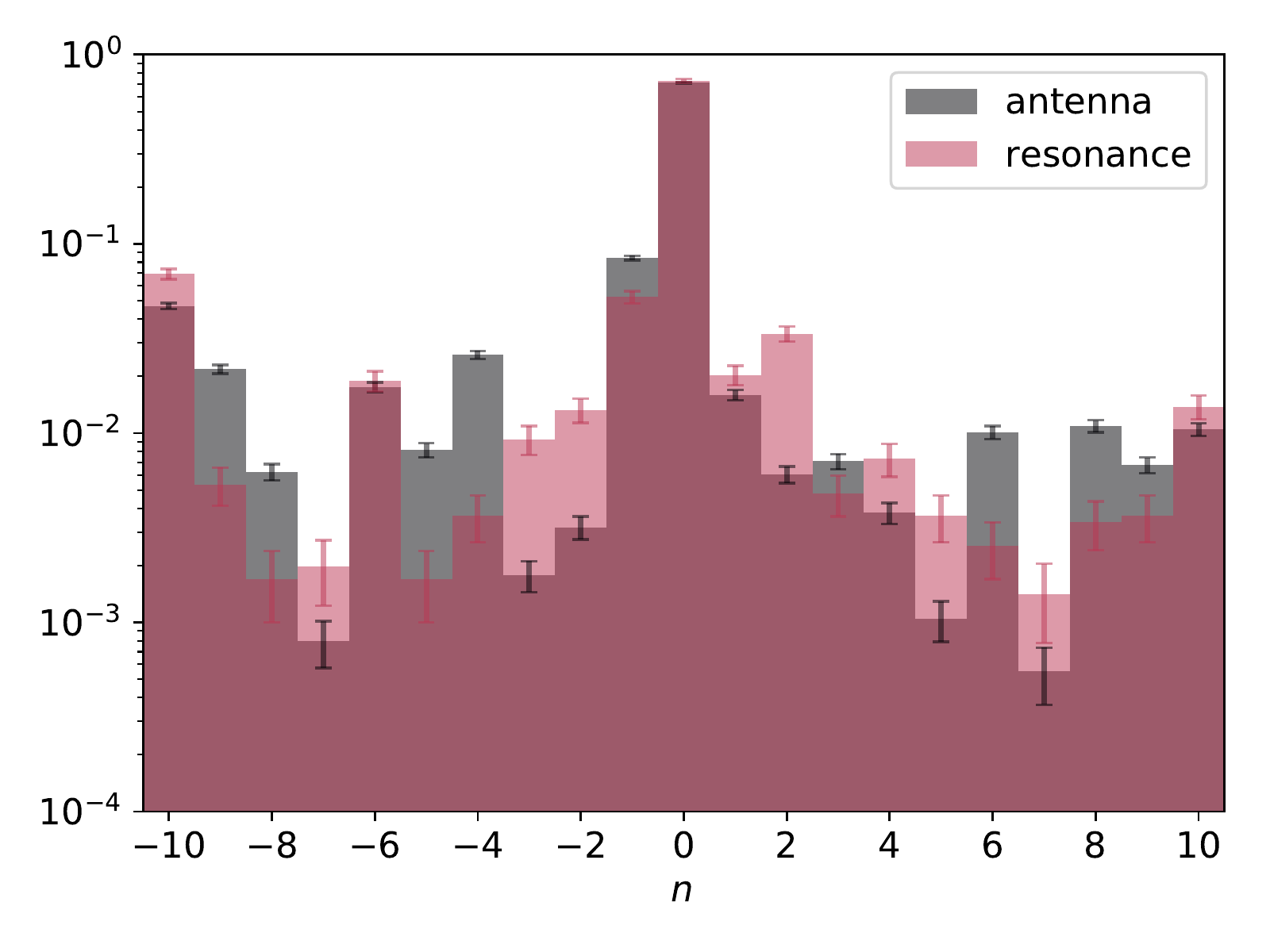}
                \caption{}
                \label{fig:nopspace}
            \end{subfigure}
            \begin{subfigure}{\halfwidth}
                \includegraphics[width=\textwidth]{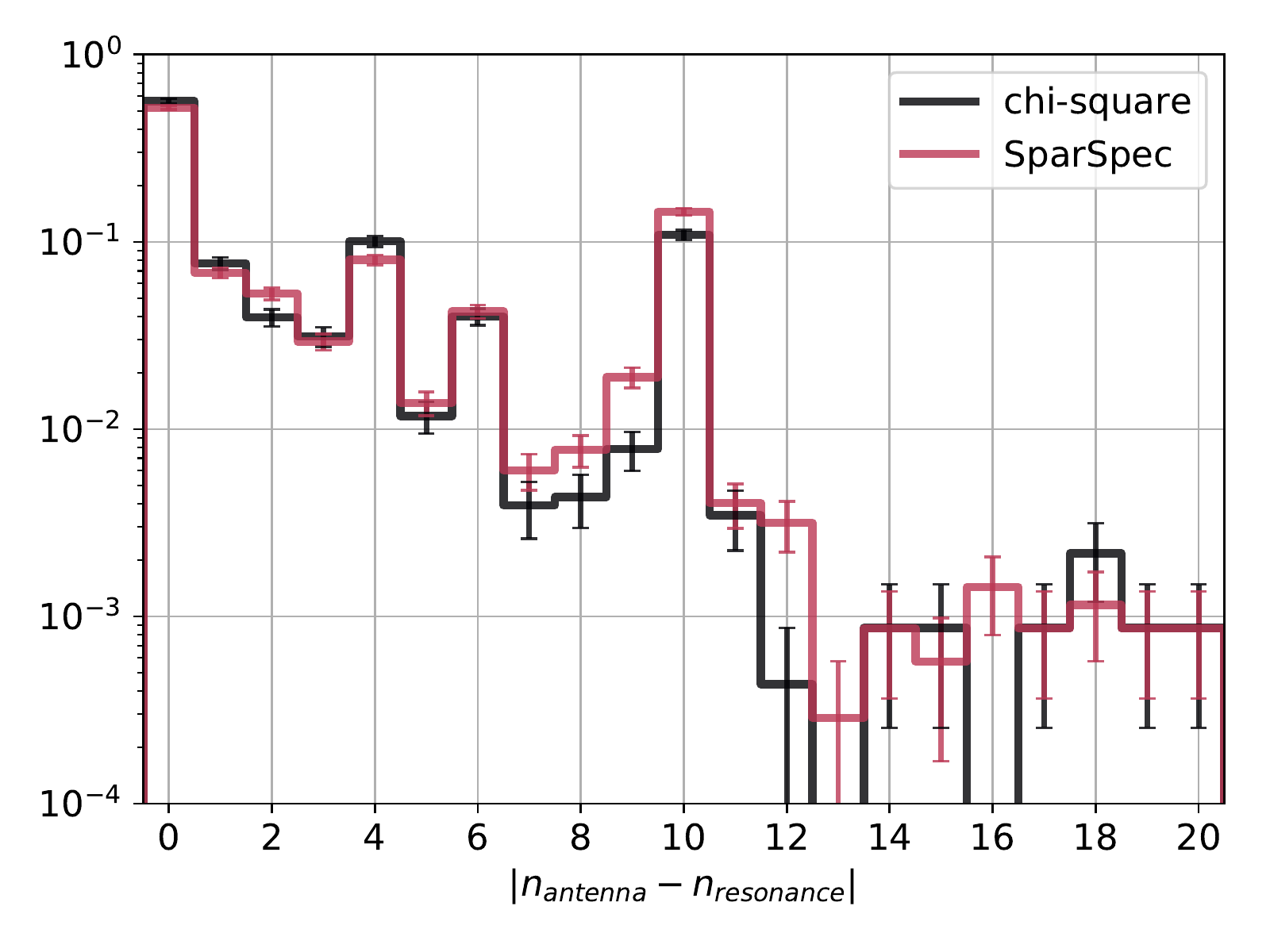}
                \caption{}
                \label{fig:absn}
            \end{subfigure}
            \caption{\blue{Histograms (normalized) of (a)~toroidal mode numbers applied by the antenna $\N{17103}$ and estimated of the resonances using \SparSpec \jm{(see \cref{sec:sparspec}, $\Ntot = 3549$)}, and (b)~the absolute difference between the applied antenna toroidal mode number and that estimated of each resonance using the chi-square \jm{(see \cref{sec:chisquare}, $\Ntot = 2328$)} and \SparSpec $\N{3505}$ methods. All data are restricted to $\absn \leq 10$, and estimations require a `confidence factor' $\cf \geq 2$ or $\cfSS \geq 2$ (see text for details). Uncertainties are shown as error bars.}}
        \end{figure}

        Following a recent system upgrade \cite{Puglia2016}, six new amplifiers allow antennas~1-5 and 7 to be powered and phased independently. This marks a significant improvement over the previous \AEAntenna feed system, which had only 0 or $\pi$ phasing. Now, antenna phases can be carefully chosen so that the injected power spectrum is maximal at toroidal mode numbers as high as $\absn \approx 20$. The operational space for the \blue{dominant} applied toroidal mode number is shown in the normalized histogram (black) of \cref{fig:nopspace}. \blue{The antenna was operated most frequently with phases $\n = 0, -1, -4$ and $-10$, with positive $\n$ defined in the direction of the plasma current $\Ip$. These were effectively randomly chosen by the operators in order to probe even vs odd and low vs high toroidal mode numbers. The predominance of negative $\n$ values in the applied mode number was an operational oversight as $\Ip$ is typically directed in the $-\phi$ direction in JET. The calculation of $\n$ will be discussed in the following section.}
        
\section{Resonance detection and parameter estimation}\label{sec:detection}

        As the antenna frequency passes through the \AE resonant frequency, the plasma responds like a driven, damped harmonic oscillator. The resulting magnetic response is measured by a set of fourteen toroidally distributed fast magnetic probes, listed in \cref{tab:probes}. This marks an improvement over past analyses for which only ten probes \blue{- or fewer -} were available \cite{Testa2014}. The magnetic signals are synchronously detected at the antenna frequency with an effective band-pass filter of width $\Delta f \approx \SI{0.1}{kHz}$ \cite{Fasoli1997}. This gives a time-evolving amplitude and phase for each probe; for example, see those in \cref{fig:res94654}. The data from all probes are then used to calculate the \AE resonant frequency $\fo$, damping rate $\gamma$, and toroidal mode number $\n$.

        \begin{table}[h!]
            \centering
            \caption{Fast magnetic probes and their toroidal positions rounded to the nearest degree. Those with names beginning with H or T are used to calculate the toroidal mode number.}
            \resizebox{\textwidth}{!}{\begin{tabular}{c | c c c c c c c c c c c c c c c c}
                \hline
                Probe & H301 & H302 & H303 & H304 & H305 & T001 & T002 & T006 & T007 & T008 & T009 & I801 & I802 & I803 \\
                Angle & 77 & 93 & 103 & 108 & 110 & 3 & 42 & 183 & 222 & 257 & 290 & 317 & 317 & 318 \\
                \hline
            \end{tabular}}
            \label{tab:probes}
        \end{table}

    \subsection{Resonant frequency and damping rate}\label{sec:freqdamprate}

\begin{figure}[h!]
            \centering
            \begin{subfigure}{\halfwidth}
                \includegraphics[width=\textwidth]{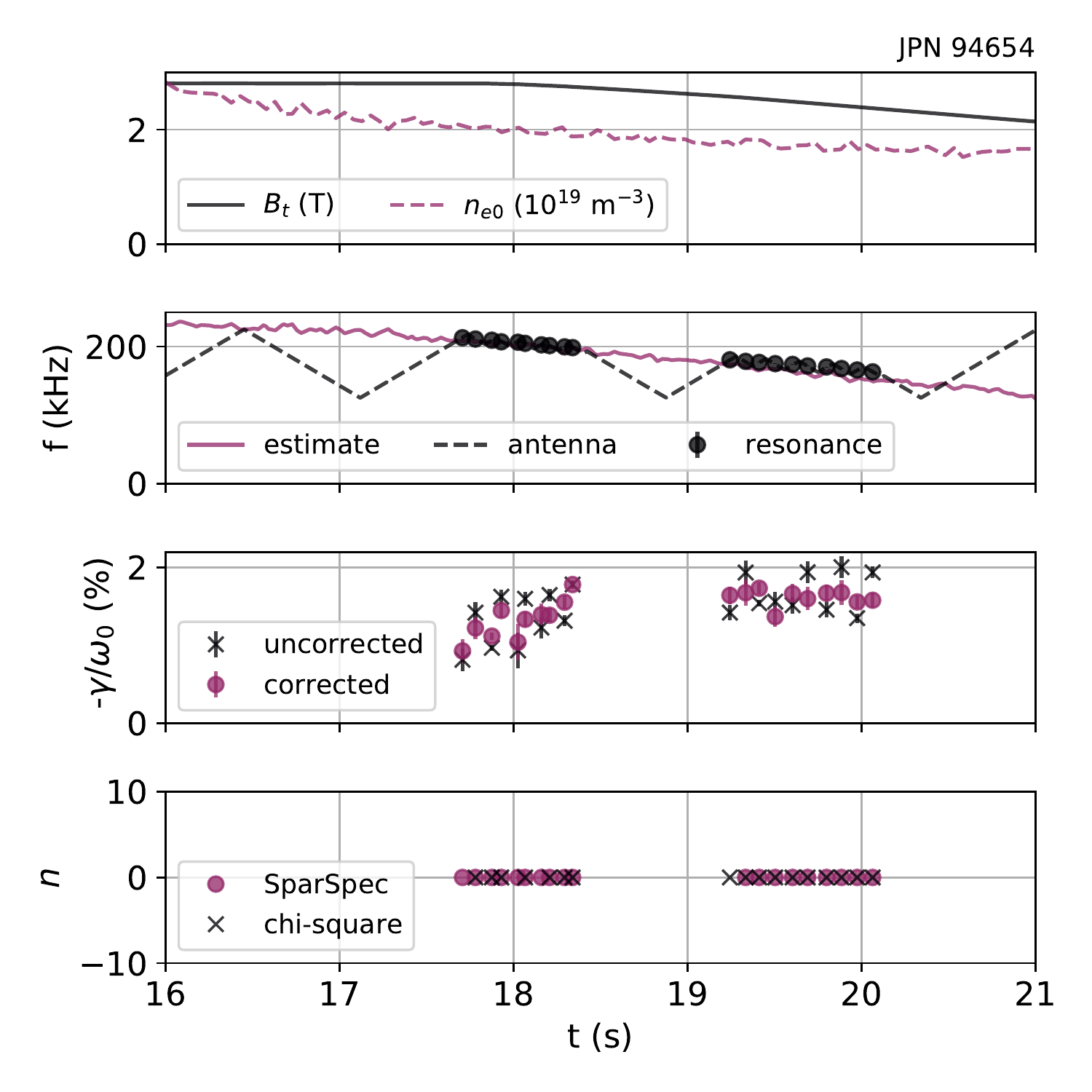}
                \caption{}
                \label{fig:params94654}
            \end{subfigure}
            \begin{subfigure}{\halfwidth}
                \includegraphics[width=\textwidth]{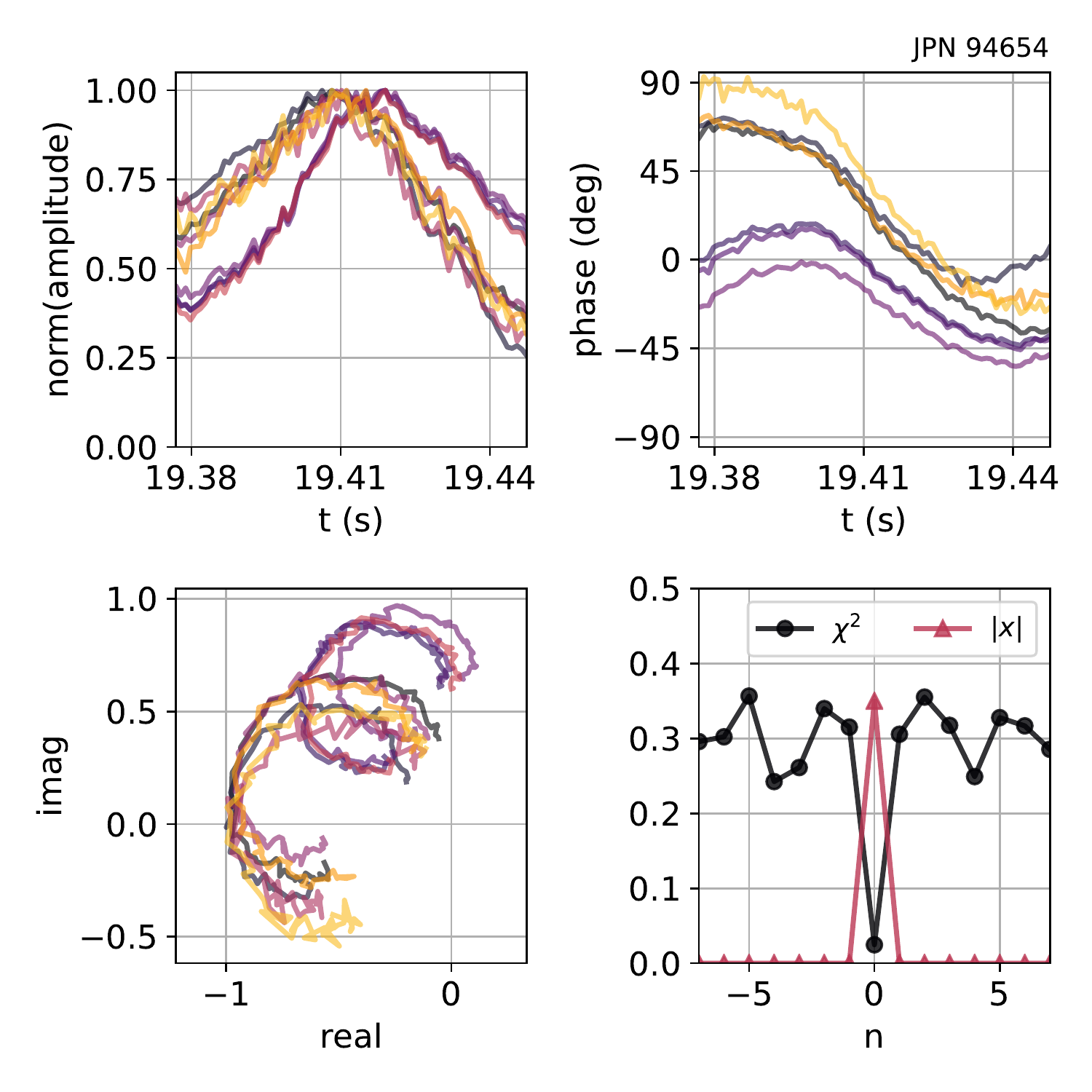}
                \caption{}
                \label{fig:res94654}
            \end{subfigure}
            \caption{(a) The toroidal magnetic field and central electron density; estimated TAE, antenna, and resonant frequencies; uncorrected and corrected damping rates; and toroidal mode numbers \blue{calculated by both \SparSpec and chi-square methods, with a  `confidence factors' $\cf \geq 2$ and $\cfSS \geq 2$} (see text for details), for JPN~94654. (b) For one resonance, data from ten fast magnetics probes: amplitudes normalized to their maxima, phases \blue{(only those used for the toroidal mode number calculation)}, complex representations, and resulting chi-square \blue{and \SparSpec amplitude spectra} limited to $\absn \leq 7$. From \cref{tab:probes}, probes used are H301-5, T006/7, and I801-3.}
        \end{figure}

        For an driven, \emph{weakly}-damped harmonic oscillator, i.e. $\abs{\go} \ll 1$, the system response to a driving frequency $\omega$ is well-approximated by the transfer function \cite{Moret1994,Fasoli1995,Fasoli1995nf}
            \begin{equation}
                H(\omega) = \frac{1}{2} \left( \frac{r}{ i(\omega - \wo) - \gamma } + \frac{r^{*}}{ i(\omega + \wo) - \gamma } \right) + \mathrm{offset},
                \label{eq:transfer}
            \end{equation}
        with $r$ a complex residue and $^{*}$ denoting the complex conjugate. The resulting pole in the complex plane can be seen in \cref{fig:res94654} for ten probes (see caption for details). A fit of \cref{eq:transfer} gives values $\wo = 2\pi\fo$ and $\go$ for each probe, along with associated uncertainties $\Delta \fo$ and $\Delta(\go)$. In this work, the final fitted values of $\fo$ and $\go$ are calculated as the mean of all probes' fits with inverse variance weighting; here, the variance is taken to be the square of the uncertainty. The total uncertainty is then calculated as the standard error of the weighted mean in a way similar to \cite{Gatz1995}, except that the inverse variance is (again) used for weighting,\footnote{As opposed to the \emph{square} of the inverse variance.} which actually makes this estimation more conservative.
        
        An automatic resonance detection algorithm was run on all 676 plasma pulses with \AEAntenna operation. \blue{Each probe was calibrated for its frequency-dependent response.} The sum of all magnetic probes' amplitudes was used to identify peaks in signal - i.e. possible resonances - in an unbiased way. Selection criteria for the data to be fit with \cref{eq:transfer} include the following: The maximum amplitude of each peak must be at least 20\% higher than its neighboring minima. The time duration of each peak must be in the range $\Delta t = \SI{10-200}{ms}$, and a phase change of $\Delta\theta = \SI{55-180}{degrees}$ must occur.\footnote{During real-time resonance tracking, the transitions between positive and negative antenna scan rates, i.e. $\rd f/\rd t \to -\rd f/\rd t$, can occur so quickly that only a small phase change, e.g. $\Delta\theta \approx \SI{1}{rad} \approx \SI{57}{degrees}$, is observed.} Any fits with uncertainties $\Delta\fo > \SI{10}{kHz}$ or $\dgo > 10\%$, or R-squared ``goodness of fit'' $\Rsquare < 0.8$ are discarded outright. Of those remaining, data from \jm{at} least three probes are needed to compute the weighted average. After this initial filter, data presented in this paper are also subject to the constraints of \cref{tab:constraints}: The first three constraints increase our confidence in the probes' collective measurement. The last constraint filters out noise due to high neutral beam injection (NBI) power and associated edge localized modes (ELMs), as done in \cite{Fasoli2010}. \blue{However, note that there are novel measurements of stable TAEs at high external heating powers (NBI + RF) $\sim \SI{25}{MW}$, which will be explored in future work.}
        
        \begin{table}[h!]
            \caption{Minimum constraints on data in this paper.}
            \centering
            \begin{tabular}{c c}
                \hline
                Parameter & Upper bound \\
                \hline
                Uncertainty in resonant frequency & $\Delta\fo \leq \SI{1}{kHz}$ \\
                Normalized damping rate & $-\go \leq 6\%$ \\
                Uncertainty in damping rate & $\dgo \leq 1\%$ \\
                NBI power & $\Pnbi \leq \SI{7}{MW}$ \\
                \hline
            \end{tabular}
            \label{tab:constraints}
        \end{table}
        
        In total, there were \blue{$\Ntot = \Ndata$}~resonances detected in \blue{479}~pulses which satisfied the above criteria. The frequencies of these resonances are shown in the histogram (purple) of \cref{fig:fopspace}. We see that the number of observations increases with frequency, with most having $\fo \geq \SI{200}{kHz}$, a typical range for \TAE frequencies in JET. An estimate of the \TAE frequency, calculated as $\ftae \approx \Bo/4\pi\qo\Ro\sqrt{\uo\meff\neo}$, is shown for pulse JPN~94654 in \cref{fig:params94654}. Here, on-axis parameters are the toroidal magnetic field $\Bo$, safety factor $\qo$, major radius $\Ro$, and electron density $\neo$; the vacuum permeability is $\uo$, and effective mass is $\meff \approx \mH(2 - n_{\H}/\ne - n_\mathrm{He3}/\ne$), with $\mH$ the mass of \hydrogen. The estimated frequency $\ftae$ and resonant frequency $\fo$ agree well for $\JPN{94654}$, and the real-time resonance tracking system is also successfully demonstrated in this pulse. 

        The ratio of fitted resonant frequencies to their corresponding estimated \TAE frequencies is shown in the histogram (black) of \cref{fig:fratio}. The histogram peaks at a ratio of $\fo/\ftae = 1.2$ and skews toward values $\fo/\ftae > 1$, which has been observed for \AEAntenna data previously \cite{Fasoli1996,Testa2004}. This can be compared with the ratio of resonant frequencies to the \TAE frequencies calculated by the MHD code \MISHKA \cite{mishka_mikhailovskii1997}, also shown in \cref{fig:fratio} (purple). To calculate $\fmishka$, the code \HELENA \cite{helena_huysmans1991} was first used to convert the magnetic geometry from \EFIT \cite{Lao1985} into the format required by code \CSCAS \cite{cscas_huysmans2001}, which calculates the \Alfven continuum. Then \MISHKA \cite{mishka_mikhailovskii1997} was used to calculate mode structures and final \TAE frequency estimates for $\n = 0 - 7$. The histogram of $\fo/\fmishka$ uses the value of $\fmishka$ with the same estimated $\absn$ as the resonance. 
        
        As expected, $\fo$ agrees better with $\fmishka$ than $\ftae$, although the histogram is still skewed toward $\fo/\fmishka > 1$. One likely cause for this discrepancy is the uncertainty in the safety factor profile calculated by standard EFIT; this can be better constrained with Motional Stark Effect or Faraday rotation data, but such data were not available for every pulse. Plasma rotation will also shift the mode frequency with respect to the lab frame; however, because rotation was not regularly diagnosed, it has not been included in this analysis.

    \subsection{Damping rate correction}\label{sec:correction}
        
        The transfer function of \cref{eq:transfer} is technically only valid for weakly damped harmonic oscillators with \emph{constant} resonant and driving frequencies. This is not the case in these experiments as both the antenna and resonant frequencies are changing in time. For most cases, $\abs{\rd \fo/ \rd t} \ll \abs{\rd \f/ \rd t}$, so this is no issue. However, when $\abs{\rd \fo/ \rd t} \sim \abs{\rd \f/ \rd t}$, the resonant peak can appear much sharper or broader than the true damping rate would allow. Modifying \cref{eq:transfer} presents a challenge as the true differential equation representing the physical system now involves additional time dependencies $\omega(t)$ and $\wo(t)$. Even linear approximations, e.g. $\omega(t) \approx \omega + \alpha t$, introduce nonlinearities which have no analytical solution.
        
        Therefore, \cref{eq:transfer} was used for the calculation of all damping rates, and a correction was applied in post-processing, as has been done previously \cite{Fasoli1997}. This corrective ``lookup table'' was assembled in the following way: The amplitude and phase of a driven, damped harmonic oscillator were simulated for a range of ``true'' damping rates and linearly varying driving and resonant frequencies, spanning all values in our database: $-\go \in [0.1\%,6\%], (\rd\fo/\rd t)/\fo \in \SI{[0,1]}{s^{-1}}$, and $(\rd\f/\rd t)/\fo \in \SI{[-1.7,1.7]}{s^{-1}}$.\footnote{For $(\rd \fo/\rd t)/\fo < 0$, the signs of $\rd \fo/\rd t$ and $\rd \f/\rd t$ can be flipped.} The resulting synthetic data were fit with \cref{eq:transfer} to create a map from ``true'' to  ``erroneous,'' or corrected to uncorrected, damping rates. Finally, the total uncertainty was taken as the sum of corrected and uncorrected uncertainties in quadrature.
        
        An example of the difference between corrected and uncorrected damping rates can be seen in \cref{fig:params94654}. In $\JPN{94654}$, the \AEAntenna scan rate was $\abs{\rd \f/\rd t} = \SI{150}{kHz/s}$, and the resonant frequency changed at a rate $-\rd \fo/\rd t \approx \SI{20-40}{kHz/s}$ as determined from real-time tracking of the mode.\footnote{For isolated resonances, i.e. no real-time tracking, $\rd \fo/\rd t$ is calculated from the estimated $\ftae$.} The uncorrected damping rate is observed to alternate between lower and higher values depending on the sign of $\rd \f/\rd t$. For most resonances, the corrected damping rate falls in between the two extremes and varies more smoothly in time. Unless otherwise noted, all damping rates reported in this paper are the corrected values, e.g. in \cref{fig:q95,fig:heating}.  

    \subsection{Toroidal mode number}\label{sec:modenumber}

        The toroidal mode number of the detected resonance is estimated using only those probes located on the outer wall at approximately the same radial and poloidal positions; these probes' names begin with H or T in \cref{tab:probes}. Of the eleven available probes, at least three must have had ``good'' resonance fits, as described above, to be added to the database; thus, there are instances of resonances for which we are confident in the fitted values of $\fo$ and $\go$, but have no estimate of $\n$.
        
        Following the standard convention \cite{Todo2019}, positive $\n$ are measured in the co-current, i.e. co-$\Ip$, direction. In JET, the normal operating scenario is $\Ip < 0$ and $\Bo < 0$, meaning both are directed clockwise ($\phi < 0$) when viewing the tokamak from above; this is the case for \emph{all} pulses in our database. Thus, positive (negative) $\n$ are oriented clockwise (counter-clockwise). As mentioned, this also explains the operational preference for negative phasing (see \cref{fig:nopspace}) which is in the $\phi > 0$ direction.
        
        \blue{The toroidal mode number is perhaps the most difficult parameter to assess of a resonance due to the reliance on multiple probes, possible superposition of modes, and aliasing effects. Yet knowing the toroidal mode number is critical to studying $\n$-dependent \AE stability. Past analyses of \AEAntenna data have used several different methods to calculate $\n$, including linear fitting \cite{Testa2003rsi} and sparse spectral decomposition \cite{Klein2008,Fasoli2010,Panis2010,Testa2010,Testa2010epl,Testa2011,Testa2011fed,Panis2012a,Panis2012b,Testa2012,Testa2014}. \jm{In the following sections,} we pursue two complementary methods of $\n$ evaluation: The first is a weighted chi-square calculation using only phase information; the second utilizes the \SparSpec algorithm \cite{Klein2008} to decompose both probe amplitude and phase information.} 
        \jm{Agreement between the methods gives us reassurance that the result is correct; disagreement provides motivation for further investigation. In the latter case, \SparSpec can help identify sub-dominant modes which may also have good - though not the best - chi-square fits.}
        
        \mysubsubsection{\blue{Chi-square evaluation}}\label{sec:chisquare}
        
        For the first method, we minimize a weighted, reduced chi-square spectrum within the range of resolvable toroidal mode numbers. For $\Np$ ``good'' probes and a range of toroidal mode numbers $\n \in [-\nmax,\nmax]$, the reduced chi-square spectrum is computed as
        \begin{equation}
            \chisquare(\n) = \frac{1}{\Np^2} \sum_{i=1}^{\Np} \left[ \sum_{j=1}^{\Np} \frac{\min\{[\n(\phi_j-\phi_i)-\theta_j]^2\}}{\sigma_j^2} \middle/ \sum_{j=1}^{\Np} \frac{1}{\sigma_j^2} \right].
            \label{eq:chisquare}
        \end{equation}
        Here, $\phi_j$ is the toroidal position of each probe $j$ (see \cref{tab:probes}), $\theta_j$ is the corresponding phase angle of the probe signal at the resonant frequency $\f = \fo$, and the inverse variance weighting uses the uncertainty of the normalized damping rate measurement $\sigma_j = \dgo$. The inner sum over all probes $j$ is the typical chi-square calculation, while the outer sum over all probes $i$ allows each probe to be considered the reference at the origin $\phi = 0$. Note that \emph{minimum} difference between angles is used in the actual computation, since $\phi$ and $\theta$ are periodic in $2\pi$.
        
        The range of resolvable toroidal mode numbers, $\absn \leq \nmax$, depends on the probes used in each evaluation of \cref{eq:chisquare}. As shown in \ref{app:nudft}, the theoretical $\nmax$ is equal to the least common denominator of all probe positions $\phi_i/\pi$, assuming that these are rational numbers \blue{and that one probe is at the origin $\phi_0 = 0$}. In practice, $\nmax$ can be computed through brute force by comparing each $\n$ of interest. In this work, we allow a generous uncertainty in the phase, $\Delta\theta = \SI{30}{degrees}$, which makes our estimate more conservative. Sometimes, $\nmax$ and $-\nmax$ are indistinguishable; in this case, the range defaults to $\n \in [-\nmax+1,\dots,\nmax]$. For this analysis, we cap the value at $\nmax \leq 10$, although the true value is often $\nmax \sim 20$ or greater. We have chosen this upper bound based on the toroidal mode numbers of \emph{destabilized} \TAEs observed in JET; for example, see those in Fig.~12 of \cite{Dumont2018}.
        
        The final estimate of the toroidal mode number $\nres$ is taken as the value which minimizes the chi-square spectrum, $\min{[\chisquare(n)}] = \chisquare(\nres)$, within a given range $\absn \leq \nmax$. To quantify our confidence in this estimate, we define a ``confidence factor'' $\cf$ as
            \begin{equation}
                \cf =  \frac{\min\left[\chisquare(\n\neq\nres)\right]}{\chisquare(\n=\nres)} \geq 1.
                \label{eq:confidence}
            \end{equation}
        In other words, the minimum $\chisquare$ value is smaller than all others in the spectrum by a factor $\cf$, and our confidence increases as $\cf$ increases.
        
        \mysubsubsection{\blue{\SparSpec evaluation}}\label{sec:sparspec}
        
        \blue{
        Borrowing a technique \cite{Bourguignon2007} from the field of astronomy, the \SparSpec code \cite{Klein2008} utilizes the ``sparse'' representation of signals - i.e. data from a limited set of unevenly spaced magnetic probes - and performs a spectral decomposition to find a superposition of toroidal modes. Details of this calculation \cite{Bourguignon2007,Klein2008}, its real-time implementation on JET \cite{Testa2010epl,Testa2011fed,Testa2014}, and associated analyses \cite{Fasoli2010,Panis2010,Testa2010,Testa2011,Panis2012a,Panis2012b,Testa2012} can be found in a variety of references. A brief overview is given here: For $N$ probes at toroidal positions $\vphi = [\phi_1,\dots,\phi_N]$, their complex-valued measurements can be represented as $\vy = [y_1,\dots,y_N]$. For a range of toroidal mode numbers $\n_j$, a matrix can be created with complex-valued components $W_{jk} = \exp(i \n_j \phi_k)$. The aim is then to minimize the function
            \begin{equation}
                J(\vx) = \norm{\vy - W \vx}^2 + \lambda \max\left( W^\dagger \vy \right) \sum_j \abs{x_j},
                \label{eq:sparspec}
            \end{equation}
        where $\lambda \in [0,1]$ is a free parameter, $W^\dagger$ is the conjugate transpose of $W$, and $x_j$ is the fitted amplitude of mode $\n_j$. When $\lambda = 0$, \cref{eq:sparspec} is just a linear least-square fit; however, for $\lambda > 0$, the second term of \cref{eq:sparspec} is a cost function penalizing additional non-zero amplitudes $x_j$.
        }
        
        \blue{
        In this work, we evaluated \SparSpec over a range of toroidal mode numbers $\absn \leq 30$ with a cost function parameter $\lambda = 0.85$, a value found to work well in previous studies \cite{Testa2010epl,Testa2011fed}. In theory, this combination should lead to noise in the signal being ``filtered out'' as low-level amplitudes at high mode numbers. Then, just as with the chi-square evaluation in the previous section, the range of toroidal mode numbers was limited to those resolvable by the available probes. In past works, this spectral decomposition was then used to compute the resonant frequency and damping rate of each individual mode contributing to the observed resonance. Here, for simplicity, we report the ``dominant'' mode $\nres$ having the largest amplitude $\abs{x_0} = \max(\abs{x_j})$. We compute another `confidence factor' $\cfSS$ similar to \cref{eq:confidence}, but comparing the maximum (absolute) amplitude to all others in the \SparSpec spectrum, i.e.
        \begin{equation}
            \cfSS = \frac{\abs{x_j(\n_j = \nres)}}{\max\abs{x_j(\n_j \neq \nres)}} \geq 1.
            \label{eq:confidenceSS}
        \end{equation}
        In other words, the absolute amplitude of the dominant mode is greater than that of each other mode by this factor $\cfSS$, and our confidence increases as $\cfSS$ increases.
        }
        
        \mysubsubsection{\blue{Results}}\label{sec:results}
        
        \blue{
        Toroidal mode number estimates using both chi-square and \SparSpec calculations, with confidence factors $\cf \geq 2$ and $\cfSS \geq 2$, respectively, are shown in \cref{fig:params94654} for $\JPN{94654}$. For this pulse, all estimates are $\nres = 0$. The chi-square and \SparSpec spectra are also shown for one resonance in \cref{fig:res94654}; both show a ``confident'' prediction of $\nres = 0$. Since \TAEs cannot have $\n=0$, this could indicate a measurement of a Global \AlfvenEigenmode (\GAE) \cite{gae_ross1982,gae_appert1982} which has been observed previously in JET; see \cite{Fasoli1997,Testa2001,Panis2012a,Oliver2017} and others.
        }
        
        \blue{
        All resonances' toroidal mode numbers, evaluated with \SparSpec and a confidence factor $\cfSS \geq 2$, are shown in the histogram (purple) of \cref{fig:nopspace}.\footnote{Note that a histogram of data from the chi-square evaluation is not shown in \cref{fig:nopspace} because it is almost identical to - i.e. agrees well with - that from \SparSpec.} As with the antenna operational space, most resonances are estimated to have $\n = 0$, with the number of observations generally decreasing as $\absn$ increases. A similar trend was observed in past \AEAntenna data; see Fig.~3 in \cite{Panis2012a}. The predominance of $\n = 0$ measurements has a few explanations: First, a subset of these could truly be \GAEs, as mentioned. Additionally, there could be a superposition of modes dominated by $\n=0$; identifying subdominant modes via \SparSpec will be explored in future work. Finally, as the number of magnetic probes with ``good'' fits decreases, the range of resolvable $\n$ often decreases as well; this biases measurements toward low-$\n$.
        } 
        
        \blue{
        The absolute difference between the applied antenna and estimated resonance mode numbers, $\abs{\nant - \nres}$, is shown in \cref{fig:absn} for both chi-square and \SparSpec evaluations with confidence factors $\cf \geq 2$ and $\cfSS \geq 2$, respectively. Importantly, \cref{fig:absn} confirms the successful operation of the \AEAntenna. Recall that the antenna and resonant toroidal mode numbers are estimated in the same way, but ultimately come from two different sources: antenna currents and magnetic signals. The histogram is peaked at a difference of zero and decreases exponentially as the separation increases. Note the ``dips'' at odd differences (i.e. $\abs{\nant - \nres} = 3, 5, \dots$) and ``peaks'' at even differences (i.e. $\abs{\nant - \nres} = 4, 6, \dots$). This is caused by the discrete antenna system injecting power into a spectrum of toroidal modes, often preserving parity; for example, see the driven $\n$-spectrum in Fig.~2 of \cite{Panis2012a}. Note that the salient peak at $\abs{\nant - \nres} = 10$ is an artifact due to the nearly $\n = 10$ spacing of a subset of probes in \cref{tab:probes} \cite{Klein2008,Fasoli2010}.
        }
        
        \blue{
        Finally, note that while we consider toroidal mode number estimations in range $\absn \leq 10$ to be most plausible, observations of $\absn > 10$ are still prevalent: $\sim 60$ measurements via the chi-square method with $\cf \geq 2$ and $\sim 200$ measurements from \SparSpec with $\cfSS \geq 3$. These will be investigated more carefully in future work.
        }


\section{Observations in plasma parameter space}\label{sec:opspace}

    In the previous section, we compared the operational space of the \AEAntenna with the resonances' parameter space. In this section, we comment on the plasma parameter space within which these resonances were observed. It is important to note that there are several layers to the exploration of this parameter space: First, there are data associated with only \emph{observations} of resonances, such as the histogram of resonant frequencies in \cref{fig:fopspace}. Then there are observational data \emph{normalized} to the antenna operational space. For example, we observed fewer resonances at low frequencies $\f = 25-\SI{50}{kHz}$, but also operated the antenna less often in that frequency range. An even deeper layer could consider the \emph{existence} of \AEs (or other resonances) in any frequency range and the required accessibility of our antenna to probe them. However, to learn this accessibility/existence space would require extensive computational efforts and would be sensitive to many uncertainties, so it is not pursued in this paper.
    
    \ppcf{Ranges of plasma parameters in our database are given in \cref{tab:params}. The 5th and 95th percentiles of each parameter distribution are denoted, meaning 5\% and 95\% of the distribution are less than these values, respectively.} These can be compared to a similar database in \cite{Panis2012a}; see Table~1 therein. Note that in \cite{Panis2012a} only ohmically heated plasmas were used to construct the database. Of the \blue{\Ndata}~resonances in our data set, the proportions measured in limiter and \xpoint magnetic configurations were $\sim$17\% and $\sim$83\%, respectively. Unless otherwise noted, data in this paper include both limiter and \xpoint configurations. The effects of plasma shaping and plasma-antenna coupling on \AE measurements have been investigated in past works \cite{Testa2001,Testa2005,Fasoli2007,Testa2010,Panis2010,Testa2011,Panis2012a,Panis2012b,Testa2014} and will be explored for our database in \ppcf{an upcoming study \cite{Tinguely2020Coupling}}. In this section, we highlight a few salient observations and trends\blue{, but note that extracting physics from the database will require careful data filtering, proper statistical analysis, and physics-based guidance}.

    \begin{table}[h!]
        \centering
        \caption{Ranges of plasma parameters for the resonance database: plasma current, on-axis toroidal magnetic field, central electron density and temperature, NBI and RF heating powers, plasma-antenna separation, ELM frequency, central and edge safety factors, edge magnetic shear, elongation, upper and lower triangularities, normalized beta and internal inductance. \ppcf{Here, 5\% of the distribution falls below the 5th percentile value; 95\% falls below than the 95th percentile.}}
        \begin{tabular}{c c c}
            \hline
            Parameter & 5th percentile & 95th percentile \\
            \hline
            $\Ip$ (MA) & 0.74 & 1.97 \\ 
            $\Bo$ (T) & 1.74 & 3.41 \\ 
            $\neo$ ($\SI{10^{19}}{m}^{-3}$) & 1.52 & 4.73 \\ 
            $\Teo$ (keV) & 1.04 & 2.50 \\ 
            $\Pnbi$ (MW) & 0.00 & 2.19 \\ 
            $\Prf$ (MW) & 0.00 & 2.86 \\ 
            $\dlcfs$ (cm) & 9.98 & 16.74 \\ 
            $\felm$ (Hz) & 0.00 & 14.30 \\ 
            $\qo$ & 0.84 & 2.21 \\ 
            $\qnf$ & 3.21 & 7.79 \\ 
            $\snf$ & 3.00 & 5.81 \\ 
            $\elon$ & 1.27 & 1.67 \\ 
            $\triu$ & 0.00 & 0.25 \\ 
            $\tril$ & 0.02 & 0.35 \\ 
            $\betaN$ & 0.10 & 0.54 \\ 
            $\li$ & 1.00 & 1.70 \\
            \hline
        \end{tabular}
        \label{tab:params}
    \end{table}

\begin{figure}[h!]
            \centering
            \begin{subfigure}{\halfwidth}
                \includegraphics[width=\textwidth]{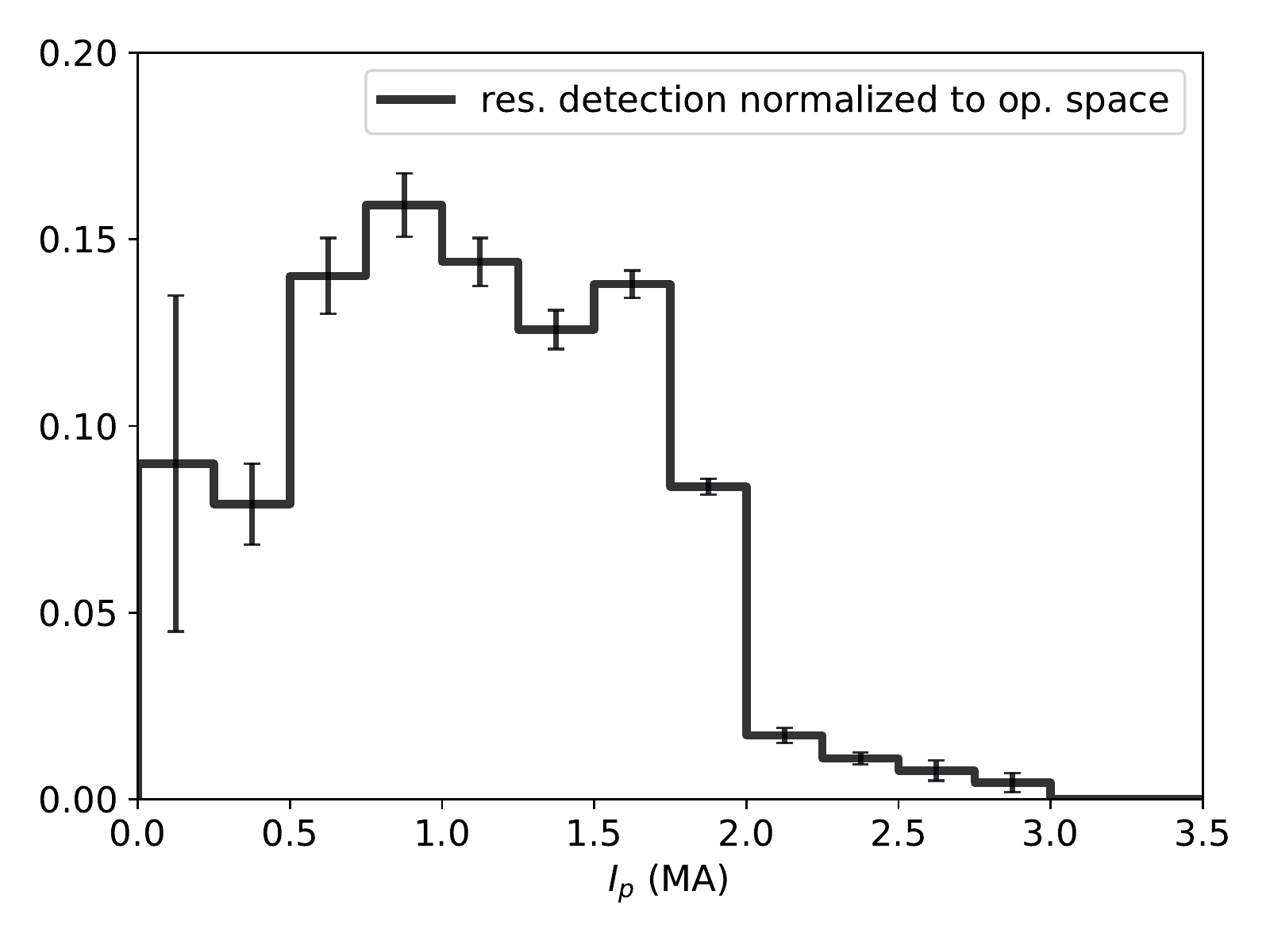}
                \caption{}
                \label{fig:Ip}
            \end{subfigure}
            \begin{subfigure}{\halfwidth}
                \includegraphics[width=\textwidth]{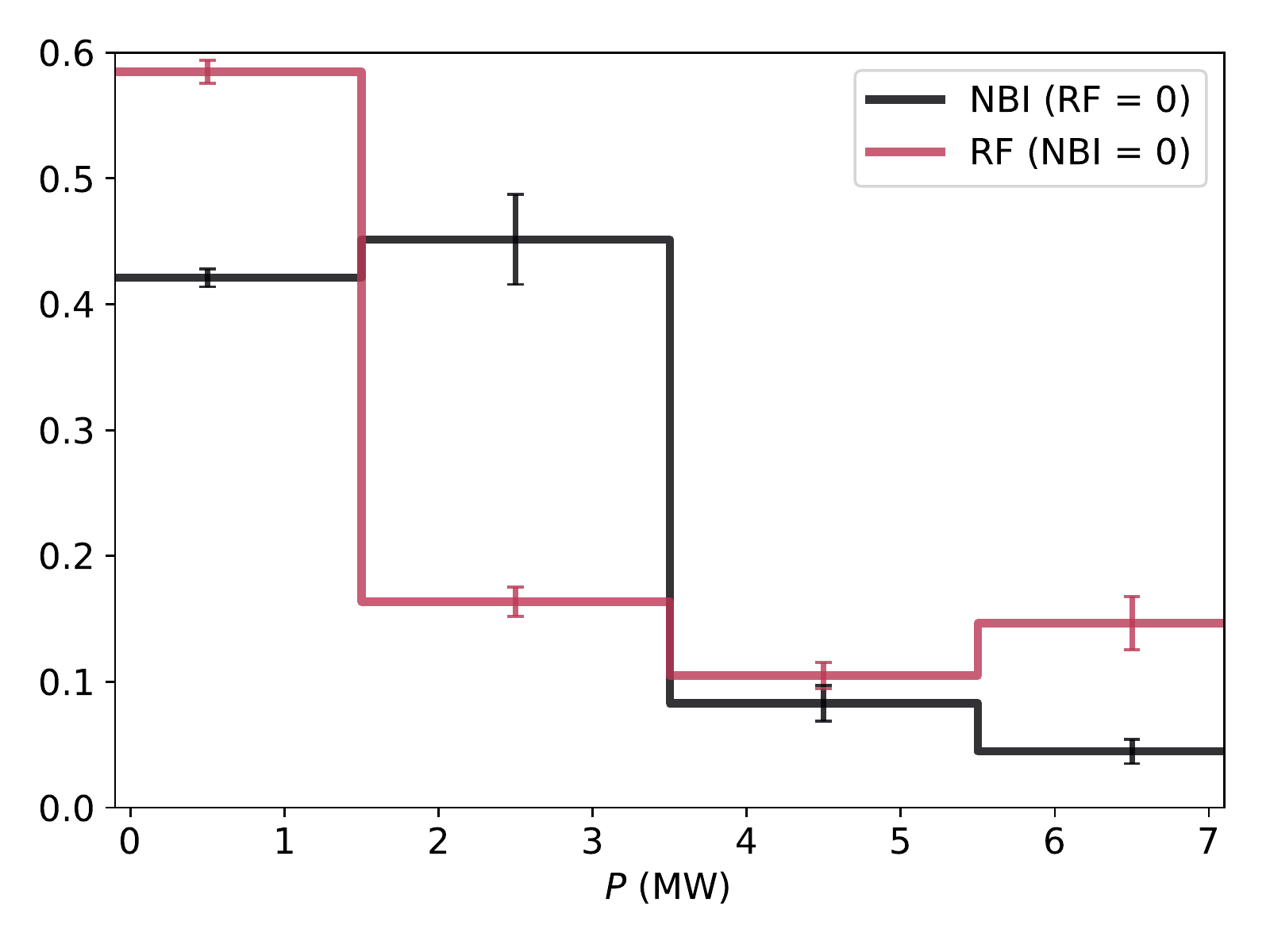}
                \caption{}
                \label{fig:heating}
            \end{subfigure}
            \caption{Histograms of the probability of resonance detection \emph{normalized to the operational space} versus (a) plasma current \blue{$\N{\Ndata}$} and \ppcf{(b) NBI \blue{$\N{3777}$} and RF \blue{$\N{4392}$} heating powers.} Uncertainties are shown as error bars.}
        \end{figure}

    The probability of resonance detection, normalized to the antenna operational space, is shown as a function of plasma current $\Ip$ in the histogram of \cref{fig:Ip}. Each bin accounts for the number of resonance observations \emph{and} the number of times the antenna operated within the bin's range. The error bars represent uncertainties from counting statistics of both values, propagated appropriately. Interestingly, there is a steep drop-off in the detection probability for plasma currents beyond $\Ip > \SI{2}{MA}$; that is, we were less likely to measure resonances when operating above $\SI{2}{MA}$. The detection probability is actually zero for $\Ip > \SI{3}{MA}$. \ppcf{One explanation for this is that the (fixed) antenna currents have a lower perturbative effect as $\Ip$ increases. The driven magnetic perturbation by the antenna at the plasma edge is of order $\delta B \approx \SI{0.1-1}{G}$ \cite{Panis2010,Puglia2016}; therefore, a threefold increase in the poloidal field strength could reduce the antenna perturbation and/or plasma response $\abs{\delta B/B}$ to below measureable levels. At the same time,} there could be a variety of other conflating factors in these high performance discharges which contribute to this observation.

    
    In \cref{fig:heating}, we show the probability of resonance detection, again normalized to the antenna operational space, as a function of heating power. \ppcf{We use wide bins, $\Delta P = \SI{2}{MW}$, in our histogram for two reasons: (i)~to improve statistics and (ii)~because external heating is usually not varied continuously, but rather operated at fixed levels.} For NBI heating only, the detection probability is relatively uniform for $\Pnbi \leq \SI{3.5}{MW}$, but drops sharply for higher powers. This could be explained by ion Landau damping from an increased population of NBI ions; such an effect has been noted before in JET \cite{Fasoli1997,Testa2003NBI,Dumont2018}. In fact, the damping rate of $\n = 1$ \TAEs was found to decrease for $\Pnbi = \SI{0-3}{MW}$ but increase beyond $\Pnbi > \SI{3}{MW}$ in \cite{Testa2003NBI}. Note that the \AEAntenna was operated for heating powers up to $\Pnbi \approx \SI{30}{MW}$ in the 2019-2020 campaign. However, as discussed previously, noise in the magnetics signals, such as that due to ELMs, is particularly prevalent for $\Pnbi > \SI{7}{MW}$ and can be misidentified as resonant peaks. Therefore, these data were excluded, as has been done in previous \AEAntenna studies \cite{Fasoli2010}. 
    
    \ppcf{
    For RF heating only, the probability of \ppcf{resonance} detection decreases sharply beyond $\Prf > \SI{1.5}{MW}$. This finding is more difficult to interpret than for only NBI heating because RF-heated fast ions can both stabilize and destabilize \AEs in JET. For example, in \cite{Fasoli1997}, increasing RF power was found to stabilize $\n = 0$ \GAEs (i.e. increase their damping rate), but destabilize $\n = 1$ \TAEs (i.e. decrease their damping rate). What can be inferred from \cref{fig:heating} is that \AE stability is more difficult to assess in high-power JET plasmas; therefore, \AEAntenna operation must be carefully optimized for success in the upcoming high-performance DT campaign. That said, the \AEAntenna should have a higher chance of success during the ``afterglow'' phase of some DT pulses, during which NBI and RF will purposefully be zeroed in order to isolate the effect of alpha drive.
    }

\begin{figure}[h!]
            \centering
            \begin{subfigure}{\halfwidth}
                \includegraphics[width=\textwidth]{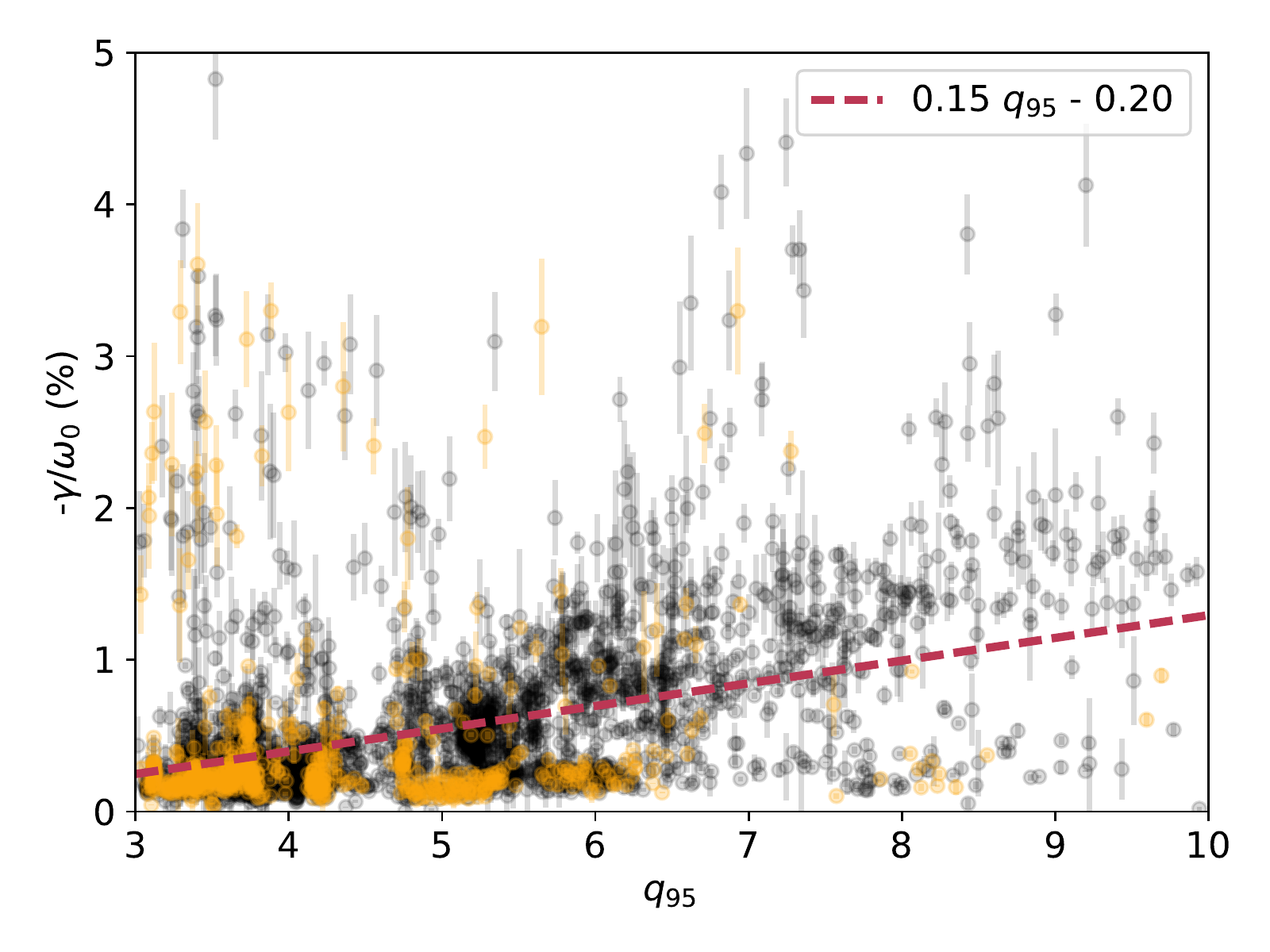}
                \caption{}
                \label{fig:q95}
            \end{subfigure}
            \begin{subfigure}{\halfwidth}
                \includegraphics[width=\textwidth]{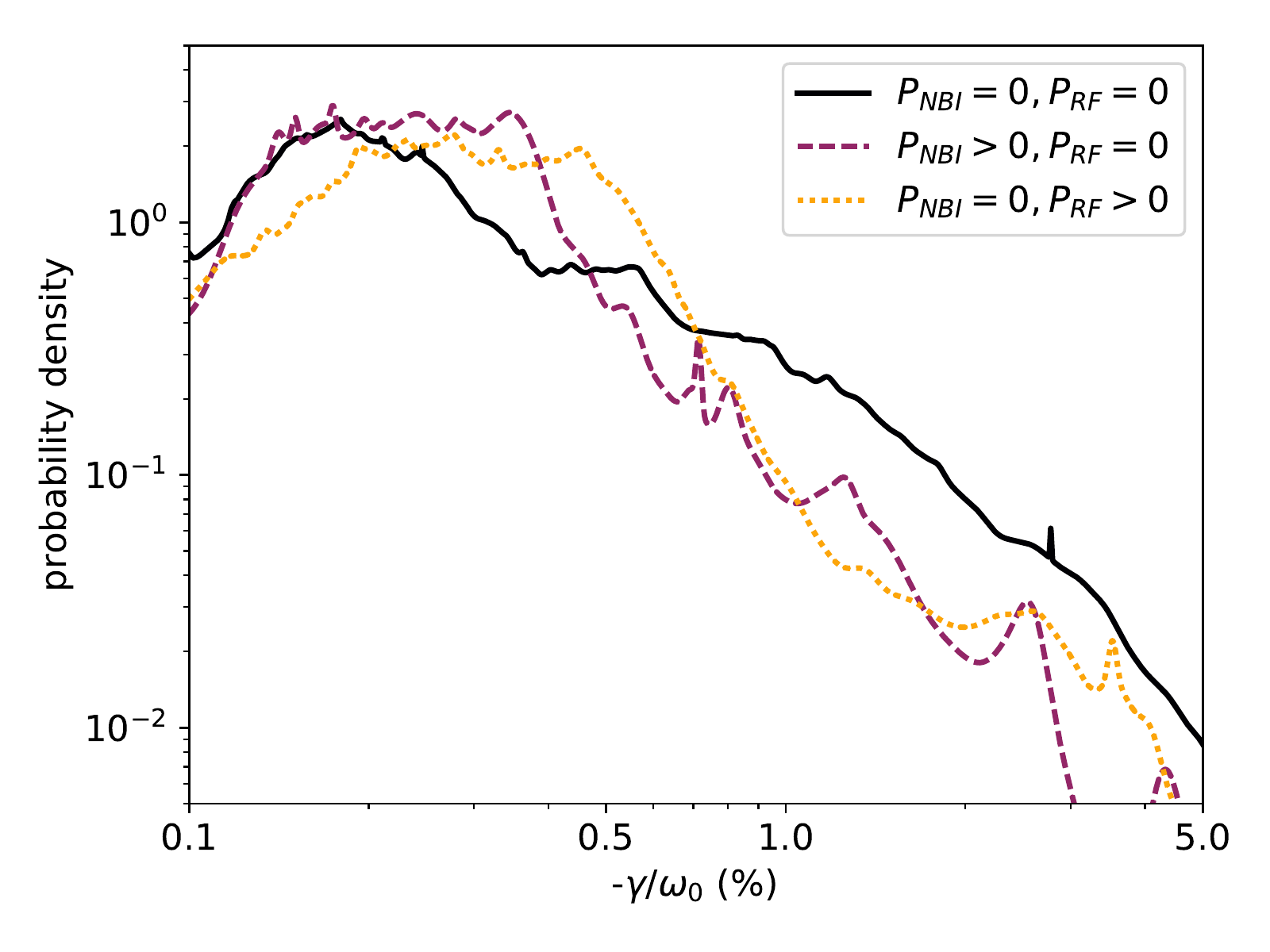}
                \caption{}
                \label{fig:pdf}
            \end{subfigure}
            \caption{\ppcf{(a) Normalized damping rate versus $\qnf$ for data with $\absn \leq 4$ (dark, $\Ntot~=~3150$) and $\absn \geq 5$ (light, $\Ntot~=~692$) in \xpoint configuration. Uncertainties are shown as error bars, and data are restricted to $\dgo \leq 0.5\%$. A linear fit to \emph{all} data is overlaid. (b) Probability density functions of the normalized damping rate during external heating from NBI only {$\N{395}$}, RF only {$\N{1025}$}, or neither {$\N{3592}$}. Note the logarithmic horizontal and vertical axes.}}
        \end{figure}

    
    \ppcf{
    Normalized damping rate measurements are shown as a function of the edge safety factor $\qnf$, as determined by \EFIT, in the scatter plot of \cref{fig:q95}. These data come only from resonances measured during \xpoint, or diverted, configuration of the magnetic geometry. Each data point is partially transparent, so that high density regions of parameter space are darker, e.g. $\qnf \in [3,6]$. Data are also distinguished by their estimated toroidal mode number: ``low'' $\absn \leq 4$ (light in color) versus ``high'' $\absn \geq 5$ (dark). Note that damping rates for data with $\absn \leq 4$ tend to be greater than those with $\absn \geq 5$. 
    }
    
    \ppcf{
    While there is significant spread in the data, we observe a general trend of increasing $\abs{\go}$ as $\qnf$ increases. This is confirmed by a simple linear fit of \emph{all} data, although the slope appears to be greater for data with $\absn \leq 4$ compared to $\absn \geq 5$. Increasing $\qnf$ - and thus changing the $q$-profile - can increase shear and continuum damping, leading to this trend. In previous studies of \AEAntenna data, the damping rate was found to increase with increasing $\qnf/\qo$ and $\qnf-\qo$ for $\absn = 3$ \TAEs \cite{Panis2012b}, but \emph{decrease} with increasing $\qnf$ for $\absn = 7$ modes \cite{Panis2012a}. The latter result is not observed in this work, but may be due to poor statistics.
    }
    
        \ppcf{
        Because the uncertainty in the damping rate can be of the same order as the damping rate itself, i.e. $\dgo \sim \abs{\go}$, it is ill-advised to visualize these data in histograms. Instead, we can construct a smooth probability density function (\pdf) in the following way: For a given data set, each measurement (data point) is assumed to be a Gaussian \pdf $g(\mu_i,\sigma_i)$ with mean $\mu_i = \go$ and standard deviation $\sigma_i = \dgo$. Then, all individual \pdfs from the data set are summed together, i.e. $G(\go) = \sum_i g(\mu_i,\sigma_i)$, and the total \pdf is normalized so that the integral is 1. The probability of a measurement within a given range is therefore just the integral of the \pdf over that range.
        } 
        
        Three \pdfs of the damping rate are shown in \cref{fig:pdf} for \ppcf{resonances detected during} no external heating, only NBI, and only RF. With no heating, the \pdf is peaked around $\abs{\go} \approx 0.2\%$ and decreases exponentially as the damping rate increases. With only NBI heating, there is an increased probability of damping rate measurements near $\abs{\go} \approx 0.3\%-0.4\%$, which could be due to NBI ion Landau damping; damping rates above $\abs{\go} > 0.5\%$ are less likely, however. A similar trend is seen for RF heating only: an increase in probability around $\abs{\go} \approx 0.3\%-0.7\%$, but a decrease beyond $\abs{\go} > 0.7\%$. As mentioned previously, this decrease in high damping rate measurements \ppcf{could} be due to \AE drive from RF-heated fast ions.


\section{Summary}\label{sec:summary}
 
    In this paper, we summarized the operation of the \AEADiagnostic, or \AEAntenna, during the 2019-2020 JET \deuterium campaign. Since its recent upgrade, six of the eight toroidally spaced antennas were independently powered and phased to excite stable MHD modes with frequencies spanning $\f = \SI{25-250}{kHz}$ (see \cref{fig:fopspace}) and toroidal mode numbers $\absn \leq \blue{30}$ (see \cref{fig:nopspace}). Synchronously detected signals from fourteen fast magnetic probes (see \cref{tab:probes}) were used to calculate mode parameters in a robust way: Resonant frequencies $\fo$ and damping rates $\gamma$ were calculated as weighted means of all (at least three) probes' individual transfer function fits (see \cref{eq:transfer} and \cref{fig:res94654}). In general, resonant frequencies agree well with both estimated \TAE frequencies and those calculated with \MISHKA, although the match is better with the latter (see \cref{fig:fratio}). The damping rate was also corrected for time-varying \AE antenna and resonant frequencies (see \cref{fig:params94654}).
    
    \blue{
    For each resonance, the toroidal mode number was estimated in two ways, via (i)~minimization of a weighted, reduced chi-square spectrum (see \cref{eq:chisquare}) and (ii)~maximization of the mode amplitude from sparse spectral decomposition (\SparSpec, see \cref{eq:sparspec}). Both methods were evaluated over the range of resolvable $n$, which depends on the positions of (at least three) probes with sufficiently good measurements of that resonance. While the discrete \AE antenna system injects power into its own $\n$-spectrum, a comparison of the dominant antenna-applied mode number and that estimated of the resonance showed good agreement (see \cref{fig:absn}). In other words, the \AEAntenna successfully excited modes with similar mode number, or at least typically preserving parity. Most common were measurements of $\n = 0$, which could be true \GAEs or caused by a superposition of modes. Observations of \TAEs generally decreased with increasing $n$ in $\absn \leq 10$ (see \cref{fig:nopspace}). However, some modes with $10 < \absn \leq 30$ were measured with high confidence ($\cf > 3$ in \cref{eq:confidence} and $\cfSS > 3$ in \cref{eq:confidenceSS}); these will be investigated in future work.
    }
    
    A database was constructed from \blue{\Ndata}~resonances detected in \blue{479}~pulses spanning a wide range of plasma parameters (see \cref{tab:params}). Data were also filtered to reduce uncertainties and remove noise (see \cref{tab:constraints}). Several initial trends were observed: The probability of resonance detection decreases sharply for plasma currents $\Ip > \SI{2}{MA}$ (see \cref{fig:Ip}); \blue{while this could simply be due to a decrease in the relative magnitude of the antenna's perturbation, there are also likely other conflating factors of high performance discharges.} Furthermore, damping rates increase with the edge safety factor (see \cref{fig:q95}), likely due to increased continuum damping. Finally, a competition between ion Landau damping and fast ion drive may be seen in two ways: First, the probability of resonance detection decreases as external heating power increases (see \cref{fig:heating}), and damping rates $-\go > 1\%$ are less likely when external heating is applied (see \cref{fig:pdf}). 
    
    This paper has laid the groundwork for many future studies utilizing this database, including statistical analyses of the bulk data as well as pulse identification for detailed analysis and comparison with modeling. Of particular interest will be the investigations high-$\n$ ($\absn \geq 7$) modes and their stability. In addition, isotope effects and, importantly, alpha drive will be explored as data is collected in the upcoming JET hydrogen, tritium, and DT campaigns.
    \jm{These data will be used to validate and improve the predictive capabilities of various MHD, kinetic, and gyrokinetic models. This is an important step in the assessment of  energetic-particle-driven \AEs and resulting \AE-enhanced transport of energetic particles in future fusion devices.}


\section*{Acknowledgments}

This work was supported by US DOE through DE-FG02-99ER54563, DE-AC05-00OR22725, and DE-AC02-05CH11231. This work has been carried out within the framework of the EUROfusion Consortium and has received funding from the Euratom research and training program 2014-2018 and 2019-2020 under grant agreement No 633053. The views and opinions expressed herein do not necessarily reflect those of the European Commission.

\appendix

\section{Calculation of the maximum resolvable toroidal mode number}\label{app:nudft}


    In this section, we will determine the range of distinguishable, or resolvable, toroidal mode numbers $\n$ for a given set of probe toroidal locations $\pk$. This is related to non-uniform/aperiodic sampling of the discrete Fourier transform.
    
    Consider a toroidal array of $\Np$ fast magnetic probes located at different toroidal angles $\pk \in [0,2\pi)$ for $k \in [1,\Np]$. For simplicity, let all probes have the same radial and poloidal position, and let $\phi_0 = 0$. For a magnetic perturbation with toroidal mode number $\n$, the phase of the (appropriately-filtered) signal of probe $k$ is $\tk = \n\pk \in [0,2\pi)$.
    
    Consider two toroidal mode numbers $\Ni$ and $\Nj$, with $\Ni > \Nj$. The signals produced by these two modes will be \emph{indistinguishable} if 
        \begin{equation}
            \mod{\Ni\pk}-\mod{\Nj\pk} = 0, \quad \forall \pk.
            \label{eq:mod}
        \end{equation}
    Here, $\mod{\cdot}$ is the modulo operator on $2\pi$. Another way to write this operator is
        \begin{equation}
            \mod{\tk} = \atantwo{\frac{\sin{\tk}}{\cos{\tk}}}
            \label{eq:atan2}
        \end{equation}
    where $\atantwo{\cdot} \in [0,2\pi)$ is the arctangent function in four quadrants. One property of this function is
        \begin{equation}
            \atantwo{\frac{\yi}{\xi}} \pm \atantwo{\frac{\yj}{\xj}} = \atantwo{\frac{\yi \xj \pm \yj \xi}{\xi \xj \mp \yi \yj}}.
        \end{equation}
    
    Let $\ti = \Ni\pk$, $\xi = \cos\ti$, $\yi = \sin\ti$, and the same for $\tj, \xj$, and  $\yj$ (where we have dropped the subscript $k$). Combining \cref{eq:mod,eq:atan2}, and using the angle summation trigonometric identities, our indistinguishability condition becomes
        \begin{equation}
            \atantwo{\frac{\sin(\ti-\tj)}{\cos(\ti-\tj)}} = \mod{(\Ni-\Nj)\pk} = 0, \quad \forall \pk.
            \label{eq:ind}
        \end{equation}
    We only need one probe location which does not satisfy \cref{eq:ind} for toroidal mode numbers $\Ni$ and $\Nj$ to be distinguishable.
    
    Presume that all $\pk$ are some rational fraction of $2\pi$.\footnote{This is not a bad assumption since there will always be some error in our measurement. Thus, we actually require that \cref{eq:ind} be less than some uncertainty in the phase, instead of exactly zero.} Then there exists $\Nstar = \Ni-\Nj$ (along with its integer multiples) which satisfies \cref{eq:ind} for all $\pk$. For a given $\Nstar$, we want to \emph{minimize} both $\abs{\Ni}$ and $\abs{\Nj}$; these are then the \emph{smallest} mode numbers which are indistinguishable. Pairs including higher values can also be indistinguishable, but are the result of aliasing. By inspection, we can minimize both $\abs{\Ni}$ and $\abs{\Nj}$ by setting $\Nj = -\Ni$. Thus, we conclude that the \emph{maximum} resolvable toroidal mode number is $\Nmax = \floor{\Nstar/2}$, with $\floor{\cdot}$ the floor operator. Note that this satisfies the Nyquist theorem for probes with uniform separation $\Delta\phi = 2\pi/\Nstar$. 
    
    It is \emph{not} always the case that both $\pm\Nmax$ can be distinguished. This must be checked separately. Hence, the range of resolvable toroidal mode number is either $\n \in [-\Nmax,\dots,\Nmax]$ or $[-\Nmax+1,\dots,\pm\Nmax]$, where $\pm\Nmax$ is treated as ``one'' toroidal mode number.
    
    To determine $\Nmax$ from $\pk$, we use the above reasoning to require that each $\pk/2\pi = \mk/2\Nmax$ is a rational number, with $\mk$ non-negative integers. (Recall that we set $\phi_0 = 0$ so that $m_0 = 0$.) Then, $\Nmax$ can be determined by finding the lowest common denominator of all $\pk/\pi$, which can be computed by various algorithms.


\section*{References}
\bibliographystyle{unsrt}
\bibliography{bib}

\end{document}